\documentclass[a4paper,UKenglish]{lipics}

\usepackage{microtype}
\usepackage{diagrams}
\usepackage{bussproofs}

\newcommand{\vect}[1]{\boldsymbol{#1}}

\newcommand{\lbi}[3]{\mathsf{let}\;#1\;\mathsf{be}\;#2\;\mathsf{in}\;#3}
\long\def\symbolfootnote[#1]#2{\begingroup%
\def\thefootnote{\fnsymbol{footnote}}\footnote[#1]{#2}\endgroup}

\newcommand{\ra}[1]{\stackrel{#1}{\longrightarrow}}

\newarrow{Squiggle}{dash}{dash}{dash}{dash}{O}
\newcommand{\sem}[2][M\!,g]{ [\![ #2 ]\!]^{}}

\newcommand{\ct}[1]{\underline{#1}}
\newcommand{\return}[0]{\mathsf{return}\;}
\newcommand{\sipm}[4]{\mathsf{pm}\;#1\;\mathsf{as}\;\langle #2,#3\rangle\;\mathsf{in}\;#4}
\newcommand{\upm}[2]{\mathsf{pm}\;#1\;\mathsf{as}\;\langle \rangle\;\mathsf{in}\;#2}
\newcommand{\force}[0]{\mathsf{force}\;}
\newcommand{\thunk}[0]{\mathsf{thunk}\;}
\newcommand{\toin}[3]{#1\;\mathsf{to}\;#2\;\mathsf{in}\;#3}

\newcommand{\Bcat}[0]{\mathcal{B}}
\newcommand{\Ccat}[0]{\mathcal{C}}
\newcommand{\Dcat}[0]{\mathcal{D}}
\newcommand{\proj}[2]{\mathbf{p}_{#1,#2}}
\newcommand{\diag}[2]{\mathsf{diag}_{#1,#2}}
\newcommand{\Set}[0]{\mathsf{Set}}
\newcommand{\diagv}[2]{\mathbf{v}_{#1,#2}}
\newcommand{\id}[0]{\mathsf{id}}
\newcommand{\Cat}[0]{\mathsf{Cat}}
\newcommand{\Id}[0]{\mathsf{Id}}
\newcommand{\self}[0]{\mathsf{self}}

\newcommand{\refl}[0]{\mathsf{refl}\;}
\newcommand{\type}[0]{\;\;\mathsf{type}}
\newcommand{\vtype}[0]{\;\;\mathsf{vtype}}
\newcommand{\ctype}[0]{\;\;\mathsf{ctype}}
\newcommand{\idpm}[3]{\mathsf{pm}\;#1\;\mathsf{as}\;(\refl #2)\; \mathsf{in}\;#3}
\newcommand{\Fam}[0]{\mathsf{Fam}}
\newcommand{\nil}[0]{\mathsf{nil\;}}
\newcommand{\tr}[0]{\mathsf{tr}\;}
\newcommand{\ctxt}[0]{\;\;\mathsf{ctxt}}
\newcommand{\print}[1]{\mathsf{print}\; #1\;.\; }
\newcommand{\diverge}[0]{\mathsf{diverge}\;}
\newcommand{\nondet}[2]{\mathsf{choose}_{#1}(#2)}
\newcommand{\readcell}[2]{\mathsf{readto}_{#1}(#2)}
\newcommand{\writecell}[1]{\mathsf{write}\;#1\;.\;}
\newcommand{\error}[1]{\mathsf{error\;#1\;}}
\newcommand{\op}[0]{\mathsf{op}}

\title{An Effectful Treatment of Dependent Types\footnote{Data types, that is. The applicability of the results of this paper to the care of individuals with a dependent personality disorder may be rather limited.}}
\titlerunning{An Effectful Treatment of Dependent Types} 
\vspace{-5pt}
\author[1]{Matthijs V\'ak\'ar}
\vspace{-5pt}
\affil[1]{University of Oxford, Dept.  Computer Science,   \texttt{matthijs.vakar@cs.ox.ac.uk}}
\authorrunning{M. V\'ak\'ar} 

\Copyright{Matthijs V\'ak\'ar}

\subjclass{F.3.2 Semantics of Programming Languages\vspace{-5pt}}
\keywords{Dependent types, effects, call-by-push-value, linear dependent types\vspace{-5pt}}

\serieslogo{}
\volumeinfo
  {Billy Editor and Bill Editors}
  {2}
  {Conference title on which this volume is based on}
  {1}
  {1}
  {1}
\EventShortName{}
\DOI{10.4230/LIPIcs.xxx.yyy.p}

\begin{document}

\maketitle
\vspace{-10pt}
\begin{abstract}We extend Levy's call-by-push-value (CBPV) analysis from simple to dependent type theory (DTT) in order to study the interaction between computational effects and dependent types. We define the naive system of dependently typed CBPV, dCBPV-, and its extension with a principle of Kleisli extensions for dependent functions, dCBPV+. We investigate these systems from the points of view of syntax, categorical semantics, concrete models and operational semantics, in presence of a range of effects. We observe that, while the expressive power of dCBPV+ is needed if we want well-defined call-by-value (CBV) and call-by-name (CBN) translations of DTT, it is a less straightforward system than dCBPV-, in presence of some effects. Indeed, to be able to construct specific models and to retain the subject reduction property in the operational semantics, we are required to impose certain subtyping conditions, the idea being that the type of a computation may only become more (not less) specified as certain effects are executed.\vspace{-5pt}
 \end{abstract}

\section{Introduction}\vspace{-5pt}
Dependent types \cite{hofmann1997syntax} are slowly being taking up by the functional programming community and are in the transition from a quirky academic hobby to a practical approach to building certified software. Purely functional dependently typed languages like Coq \cite{Coq:manual} and Agda \cite{norell2007towards} have existed for a long time. If the technology is to become more widely used in practice, however, it is crucial that dependent types can be smoothly combined with the wide range of effects that programmers make use of in their day to day work, like non-termination and recursion, mutable state, input and output, non-determinism, probability and non-local~control.

Although some languages exist which combine dependent types and effects, like Cayenne \cite{augustsson1998cayenne}, $\Pi\Sigma$ \cite{altenkirch2010pisigma}, Zombie \cite{casinghino2014combining}, Idris \cite{brady2013idris}, Dependent ML \cite{xi1999dependent} and F$\star$ \cite{swamy2015dependent}, there have always been some strict limitations. For instance, the first four only combine dependent types with first class recursion (although Idris has good support for emulating other effects), Dependent ML constrains types to depend only on static natural numbers and F$\star$ does not allow types to depend on effectful terms at all (including non-termination). Somewhat different is Hoare Type Theory (HTT) \cite{nanevski2006polymorphism}, which defines a programming language for writing effectful programs as well as a separation logic encoded in a system of dependent types for reasoning about these programs. We note that the programming fragment is not merely an extension of the logical one, which would be the elegant solution suggested by the Curry-Howard~correspondence.

The sentiment of most papers discussing the marriage of these ideas seems to be that dependent types and effects form a difficult though not impossible combination. However, as far as we are aware, treatment has so far been on a case-by-case basis and no general theoretical analysis has been given which discusses, on a conceptual level, the possibilities, difficulties and impossibilities of combining general computational effects and dependent~types.

In a somewhat different vein, there has long been an interest in combining linearity and dependent types. This was first done in Cervesato and Pfenning's LLF \cite{cervesato1996linear}. Recently, a semantic analysis of LLF was given by the author in \cite{vakar2014syntax,vakar2015syntax} which has  proved important e.g. in the development of a game semantics for dependent types. One aspect that this abstract semantics as well as the study of particular models highlight is - more so than in the simply typed case - the added  insight and flexibility obtained by decomposing the $!$-comonad into an adjunction\footnote{Indeed, connectives seem to be most naturally formulated one either the linear or cartesian side: $\Sigma$- and $\Id$-constructors operate on cartesian types while $\Pi$-constructors operate on linear types.}. This corresponds to working with dependently typed version of Benton's LNL-calculus \cite{benton1995mixed} rather than Barber and Plotkin's DILL \cite{barber1996dual}, as was done in \cite{krishnaswami2015integrating}.

Similarly, it has proved problematic to give a dependently typed version of Moggi's monadic metalanguage \cite{moggi1991notions}. We hope that this paper illustrates that also in this case a decomposed adjunction perspective, like that of CBPV \cite{levy2012call}, is more flexible than a monadic perspective. Recall from \cite{benton1996linear} that if we decompose both linear logic and the monadic metalanguage into an adjunction, we can see the former to be a restricted case of the latter which only describes (certain) commutative effects.

In this paper, we show that the analysis of linear DTT of \cite{vakar2015syntax,krishnaswami2015integrating,vakar2014syntax} generalises straightforwardly to general (non-commutative) effects to give a dependently typed CBPV calculus that we call dCBPV-, which allows types to depend on values (including thunks of computations) but which lacks a Kleisli extension (or sequencing) principle for dependent functions. This calculus is closely related to Harper and Licata's dependently typed polarized intuitionistic logic \cite{licata2009positively}. Its categorical semantics is obtained from that for linear dependent types, by relaxing a condition on the adjunction which would normally imply, among other things, the commutativity of the effects described. It straightforwardly generalises Levy's adjunction models for CBPV \cite{levy2005adjunction} (from locally indexed categories to more general comprehension categories \cite{jacobs1993comprehension}) and, in a way, simplifies Moggi's strong monad models for the monadic metalanguage \cite{moggi1991notions}, as was already anticipated by Plotkin in the late 80s: in a dependently typed setting the monad strength follows straightforwardly from the natural demand that its adjunction is compatible with substitution and, similarly, the distributivity of coproducts follows from their compatibility with substitution. In fact, we believe the categorical semantics of CBPV is most naturally understood as a special case of a that of dCBPV-. Particular examples of models are given by models of linear DTT and by Eilenberg-Moore adjunctions for strict\footnote{For brevity, from now on we shall drop the modifier "strict" for indexed structures. For instance, if we mention an indexed honey badger, we shall really mean a strict indexed honey badger.} indexed monads on models of pure DTT. The small-step operational semantics for CBPV of \cite{levy2012call} transfers to dCBPV- without any difficulties with the expected subject reduction and (depending on the effects considered) strong normalization and determinacy results.

When formulating  candidate CBV- and CBN-translations of DTT into dCBPV-, it becomes apparent that the latter is only well-defined if we work with the weak (non-dependent) elimination rules for positive connectives, while the former is ill-defined altogether. To obtain a CBV-translation and the CBN-translation in all its glory, we have to add a principle of Kleisli extensions (or sequencing) for dependent functions to dCBPV-. We call the resulting calculus dCBPV+, to which we can easily extend our categorical and operational semantics. Normalization and determinacy results for the operational semantics remain the same. However, depending on the effects we consider, we may need to add extra coercion rules to the calculus to salvage subject reduction. These embody the idea that a computation can only obtain a more (not less) precise type after certain effects, like non-deterministic branching, are executed. We analyse on a case-by-case basis the principle of dependent Kleisli extensions in dCBPV- models of a range of effects. This leads us to the same subtyping~conditions.

Before concluding, we discuss of the possibility of adding some additional connectives. In particular, we address the matter of a dependently typed enriched effect calculus (EEC) \cite{egger2009enriching}.

One the one hand, we hope this analysis gives a helpful theoretical framework in which we can study various combinations of dependent types and effects from an algebraic, denotational and operational point of view. It gives a robust motivation for the equations we should expect to hold in both CBV- and CBN-versions of effectful DTT, through their translations into dCBPV, and it guides us in modelling dependent types in effectful settings like game semantics. Moreover, it explains why combinations of dependent types and effects are slightly more straightforward in CBN than in CBV, as dependent Kleisli extensions are not required.

On the other, noting that not all effects correspond to sound logical principles, an~ expressive system like CBPV or a monadic language, with fine control over where effects occur,~is~an excellent combination with dependent types as it allows us to  use the language both for writing effectful programs and pure logical proofs about these programs. Similar to HTT in aim, but different in  implementation, we hope that dCBPV can be expanded in future to an elegant language, serving both for writing effectful programs and for reasoning about~them.

\vspace{-5pt}
\section{A Very Brief Recap of Simple Call-by-Push-Value}
\vspace{-5pt}
We briefly recall the spirit of Levy's CBPV analysis of simple type theory \cite{levy2012call,levy2006call,levy2005adjunction}. The long version of this paper \cite{vakar2015framework} includes a detailed exposition, aimed at generalising~smoothly to dependent types. CBPV roughly gives an adjunction decomposition of Moggi's monadic metalanguage \cite{moggi1991notions}. This allows one to not only fully and faithfully encode Moggi's metalanguage and with that CBV $\lambda$-calculi, but also CBN ones. CBPV has a natural small-step operational semantics, satisfying subject reduction and, depending on the effects considered, determinism at each step and strong normalization. For terms of ground type, this reproduces the usual CBV and CBN operational semantics in presence of a wide range of effects.

CBPV makes a clear distinction between the worlds of \emph{value types} (inhabited by \emph{values}) and \emph{computation types} (the home of \emph{stacks} and, in particular, \emph{computations}), between which we have an adjunction of half-modalities $F\dashv U$, which yield as special cases Moggi's modality $T=UF$ and the linear logic exponential $!=FU$. $FA$ can be read as the type of computations that can \emph{return} values of type $A$ and $U\ct{B}$ as the type of \emph{thunks} of computations of type $\ct{B}$. By only including positive type formers like inductive types on value types and negative ones like coinductive types on computation types, one is able to retain the entire $\beta\eta$-equational theory even in the presence of effects. As a consequence, we obtain an elegant categorical semantics in terms of adjunctions of locally indexed~categories. 

\vspace{-0pt}
\section{A Syntax for Dependently Typed Call-by-Push-Value (dCBPV)}\label{sec:syn}
\vspace{-5pt}
We generalise CBPV, allowing types to depend on values, but not computations (cf.~linear \cite{cervesato1996linear} or polarised \cite{licata2009positively} DTT, where types can only depend on cartesian or positive terms,~respectively). dCBPV makes~the~following judgements: well-formedness of contexts $\vdash\Gamma;\Delta\ctxt$, where $\Gamma$ is a list of distinct identifier declarations of value types and $\Delta$ is a list of  declarations of computation type (writing $\cdot$ for the empty list and $\Gamma$ as a shorthand for $\Gamma;\cdot$),~well-formedness of value types $\Gamma\vdash A\vtype$ and computation types $\Gamma\vdash \ct{B}\ctype$ (sometimes underlined to be explicit), typing of values $\Gamma\vdash^v V:A$, computations $\Gamma\vdash^c M:\ct{B}$ and~stacks $\Gamma;\Delta\vdash^k K:\ct{B}$, equality judgements for contexts, types (in context) and (typed) terms (in context). 

\nopagebreak
As types can depend on terms in a dependently typed system, we define both in one inductive definition. A discussion of the syntactic subtleties of DTT can be found in \cite{hofmann1997syntax}.

To start with, we have rules, which we shall not list, stating that all judgemental equalities are equivalence relations and that all term, type and context constructors as well as substitutions respect judgemental equality. In similar vein, we have conversion rules, stating that we may swap contexts and types for judgementally equal ones in all judgements. To form contexts, we have the rules of figure \ref{fig:ctxtrules} and, to form types, those of figure \ref{fig:depcbpvtypes}.\vspace{-5pt}
\begin{figure}[!ht]
\fbox{\resizebox{\linewidth}{!}{
\parbox{1.17\linewidth}{
\begin{tabular}{lll}
\AxiomC{}
\UnaryInfC{$\cdot;\cdot \ctxt$}
\DisplayProof
& \hspace{50pt}
\AxiomC{$\vdash\Gamma;\Delta\ctxt$}
\AxiomC{$\Gamma\vdash A\;\vtype$}
\BinaryInfC{$\vdash \Gamma,x:A;\Delta \ctxt$}
\DisplayProof

&
\hspace{50pt}
\AxiomC{$\vdash \Gamma;\cdot\ctxt$}
\AxiomC{$\Gamma \vdash \ct{B}\ctype$}
\BinaryInfC{$\vdash \Gamma;\ct{B}\ctxt$}
\DisplayProof\\
\end{tabular}
}
}}
\caption{\label{fig:ctxtrules} Rules for contexts of dCBPV, where $x$ is assumed to be a fresh identifier.}
\end{figure}\vspace{-15pt}\nopagebreak
\begin{figure}[!ht]
\fbox{\resizebox{\linewidth}{!}{
\parbox{1.17\linewidth}{
\begin{tabular}{llll}
\AxiomC{$\Gamma,x:A,\Gamma'\vdash A'\vtype$}
\AxiomC{$\Gamma\vdash^v V:A$}
\BinaryInfC{$\Gamma,\Gamma'[V/x]\vdash A'[V/x]\vtype$}
\DisplayProof
&
\AxiomC{$\Gamma\vdash \ct{B}\ctype$}
\UnaryInfC{$\Gamma\vdash U\ct{B}\vtype$}
\DisplayProof
& &\\
&&&\\\AxiomC{$\Gamma\vdash^v V:A$}
\AxiomC{$\Gamma\vdash^v V':A$}
\BinaryInfC{$\Gamma\vdash \Id_A(V,V')\vtype$}
\DisplayProof
&
\AxiomC{$\vdash \Gamma\ctxt$}
\UnaryInfC{$\Gamma\vdash 1\vtype$}
\DisplayProof
&
\AxiomC{$\{\Gamma\vdash A_i\vtype\}_{1\leq i \leq n}$}
\UnaryInfC{$\Gamma\vdash \Sigma_{1\leq i\leq n}A_i\vtype$}
\DisplayProof
&
\AxiomC{$\Gamma,x:A\vdash A'\vtype$}
\UnaryInfC{$\Gamma\vdash \Sigma_{x:A}A'\vtype$}
\DisplayProof
\\
&&\\
\AxiomC{$\Gamma,x:A,\Gamma'\vdash \ct{B}\ctype$}
\AxiomC{$\Gamma\vdash^v V:A$}
\BinaryInfC{$\Gamma,\Gamma'[V/x]\vdash \ct{B}[V/x]\ctype$}
\DisplayProof
& 
\AxiomC{$\Gamma\vdash A\vtype$}
\UnaryInfC{$\Gamma\vdash FA\ctype$}
\DisplayProof &

\AxiomC{$\{\Gamma\vdash \ct{B_i}\ctype\}_{1\leq i \leq n}$}
\UnaryInfC{$\Gamma\vdash \Pi_{1\leq i\leq n}\ct{B_i}\ctype$}
\DisplayProof &
\AxiomC{$\Gamma,x:A\vdash \ct{B}\ctype$}
\UnaryInfC{$\Gamma\vdash \Pi_{x:A}\ct{B}\ctype$}
\DisplayProof
\end{tabular}}
}
}
\caption{\label{fig:depcbpvtypes} Rules for forming the types of dCBPV.}
\end{figure}\vspace{-5pt}\\
For these types, we consider the values and computations formed using the rules of figure \ref{fig:vcdepterms}, to which we could add the obvious admissible weakening and exchange rules.
\begin{figure}[!ht]
\centering
\fbox{\resizebox{\linewidth}{!}{\parbox{1.17\linewidth}{
\begin{tabular}{llll}
\AxiomC{$\vdash \Gamma,x:A,\Gamma'\ctxt$}
\UnaryInfC{$\Gamma,x:A,\Gamma'\vdash^v x:A$}
\DisplayProof & & 
\AxiomC{$\Gamma\vdash^v V:A$}
\AxiomC{$\Gamma,x:A,\Gamma'\vdash^{v/c} R:{B}$}
\BinaryInfC{$\Gamma,\Gamma'[V/x]\vdash^{v/c} \lbi{x}{V}{R} :{B}[V/x]$}
\DisplayProof\hspace{-400pt}\;
&\\
&&&\\
\AxiomC{$\Gamma,z:UF A,\Gamma'\vdash \ct{B}\ctype$}
\AxiomC{$\Gamma\vdash^c M:FA$}
\AxiomC{$\Gamma,x:A,\Gamma'[\tr x/z]\vdash^c N:\ct{B}[\tr x/z]$}
\TrinaryInfC{$\Gamma,\Gamma'[\thunk M/z]\vdash^c \toin{M}{x}{N} : \ct{B}[\thunk M/z]$.}
\DisplayProof\hspace{-400pt}\;\\
&&&\\
\AxiomC{$\Gamma\vdash^c M:\ct{B}$}
\UnaryInfC{$\Gamma\vdash^v \thunk M:U\ct{B}$}
\DisplayProof
&
\AxiomC{$\Gamma\vdash^v V: U\ct{B}$}
\UnaryInfC{$\Gamma\vdash^c \force V: \ct{B}$}
\DisplayProof &
\AxiomC{$\vdash \Gamma\ctxt$}
\UnaryInfC{$\Gamma\vdash^v\langle\rangle :1$}
\DisplayProof &
\AxiomC{$\Gamma\vdash^v V:1$}
\AxiomC{$\Gamma\vdash^{v/c} R:{B}[\langle\rangle/z]$}
\BinaryInfC{$\Gamma\vdash^{v/c} \upm{V}{R}:{B}[V/z]$}
\DisplayProof\\
&&&\\
\AxiomC{$\Gamma\vdash^v V_i: A_i$}
\UnaryInfC{$\Gamma\vdash^v \langle i,V_i\rangle : \Sigma_{1\leq i\leq n}A_i$}
\DisplayProof \hspace{-400pt}\;
&&
\AxiomC{$\Gamma\vdash^v V: \Sigma_{1\leq i\leq n}A_i$}
\AxiomC{$\{\Gamma,x:A_i\vdash^{v/c} R_i : {B}[\langle i,x\rangle/z]\}_{1\leq i\leq n}$}
\BinaryInfC{$\Gamma\vdash^{v/c} \sipm{V}{i}{x}{R_i} : {B}[V/z]$}
\DisplayProof\hspace{-400pt}\; &\\
&&&\\
\AxiomC{$\Gamma\vdash^v V_1:A_1$}
\AxiomC{$\Gamma\vdash^v V_2:A_2[V_1/x]$}
\BinaryInfC{$\Gamma\vdash^v \langle V_1,V_2\rangle :\Sigma_{x:A_1} A_2$}
\DisplayProof\hspace{-400pt}\;
&&
\AxiomC{$\Gamma\vdash^v V: \Sigma_{x:A_1} A_2$}
\AxiomC{$\Gamma,x:A_1,y:A_2\vdash^{v/c} R:{B}[\langle x,y\rangle/z]$}
\BinaryInfC{$\Gamma\vdash^{v/c} \sipm{V}{x}{y}{R}:{B}[V/z]$}
\DisplayProof\hspace{-400pt}\;&\\
&&&\\
\AxiomC{$\Gamma\vdash^v V:A$}
\UnaryInfC{$\Gamma\vdash^v \refl{V}:\Id_A(V,V)$}
\DisplayProof \hspace{-400pt}\;& &
\AxiomC{$\Gamma\vdash^v V:\Id_A(V_1,V_2)$}
\AxiomC{$ \Gamma,x:A\vdash^{v/c} R :B[x/x',\refl{x}/p]$}
\BinaryInfC{$\Gamma\vdash^{v/c} \idpm{V}{x}{R}:B[V_1/x,V_2/x',V/p]$}
\DisplayProof \hspace{-400pt}\; &\\
&&&\\
\AxiomC{$\{\Gamma\vdash^c M_i :\ct{B_i}\}_{1\leq i\leq n}$}
\UnaryInfC{$\Gamma\vdash^c \lambda_i M_i : \Pi_{1\leq i\leq n}\ct{B_i}$}
\DisplayProof
&
\AxiomC{$\Gamma\vdash^c M: \Pi_{1\leq i\leq n}\ct{B_i}$}
\UnaryInfC{$\Gamma\vdash^c i\textquoteleft M : \ct{B_i}$}
\DisplayProof &
\AxiomC{$\Gamma,x:A\vdash^c M:\ct{B}$}
\UnaryInfC{$\Gamma\vdash^c \lambda_xM:\Pi_{x:A}\ct{B}$}
\DisplayProof
&
\AxiomC{$\Gamma\vdash^v V:A$}
\AxiomC{$\Gamma\vdash^c M:\Pi_{x:A}\ct{B}$}
\BinaryInfC{$\Gamma\vdash^c V\textquoteleft M : \ct{B}[V/x]$}
\DisplayProof
\end{tabular}
}}
}
\caption{\label{fig:vcdepterms} Values and computations of dCBPV+. Those of dCBPV- are obtained by demanding the side condition that $z$ is not free in $\Gamma';\ct{B}$ in the rule for forming $\toin{M}{x}{N}$. A rule involving $\vdash^{v/c}$ is a
shorthand for two rules: one with $\vdash^{v}$ and one with $\vdash^{c}$ in both the hypothesis and conclusion. Indices $i$ are a slight abuse of notation: e.g. $\lambda_i M_i$ is an abbreviation for~$\lambda_{(1,\ldots,n)}(M_1,\ldots,M_n)$.\vspace{-5pt}}
\end{figure}\\
As anticipated by Levy \cite{levy2012call}, the only rule that requires care is that for forming a sequenced computation $\toin{M}{x}{N}$. He suggested that $z$ should not be free in the context $\Gamma';\ct{B}$. We call the resulting system with this restriction dCBPV- and the more permissive system where we drop this restriction and allow \emph{Kleisli extensions of dependent functions} dCBPV+. We leave the discussion of stacks until section \ref{sec:op}.

We generate judgemental equalities for values and computations through the rules of figure \ref{fig:eqs}. Note that we are using extensional $\Id$-types, in the sense of $\Id$-types with an $\eta$-rule. This is only done for the aesthetics of the categorical semantics. They may not be suitable for an implementation, however, as they make type checking undecidable for the usual reasons \cite{hofmann1997syntax}. The syntax and semantics can just as easily be adapted to intensional $\Id$-types, the obvious choice to for an implementation.

\begin{figure}[!ht]
\fbox{\hspace{-5pt}
\resizebox{\linewidth}{!}{
\parbox{1.17\linewidth}
{
\begin{tabular}{ll}
$\toin{(\return V)}{x}{M} = M[V/x]$ & $M =\toin{M}{x}{\return x}$\\
$\force\thunk M=M$ & $V=\thunk\force V$\\
$\sipm{\langle i, V\rangle }{i}{x}{R_i}=R_i[V/x]$ & $R[V/z]\stackrel{\#x}{=}\sipm{V}{i}{x}{R[\langle i,x\rangle/z]}$\\
$\upm{\langle\rangle}{R}=R$ & $R[V/z]=\upm{V}{R[\langle\rangle/z]}$\\
$\sipm{\langle V,V'\rangle }{x}{y}{R}=R[V/x,V'/y]$ \hspace{20pt}\;& $R[V/z]\stackrel{\#x,y}{=}\sipm{V}{x}{y}{R[\langle x,y\rangle/z]}$\\
$\idpm{(\refl V)}{x}{R} = R[V/x]$ & $R[V_1/x,V_2/y,V/z] \stackrel{\#w}{=} \idpm{V}{w}{R[w/x,w/y,(\refl w)/z]}$\\
$i\textquoteleft \lambda_j M_j=M_i$ & $M=\lambda_i i\textquoteleft M$\\
$V\textquoteleft \lambda_x M=M[V/x]$ & $M\stackrel{\#x}{=}\lambda_x x\textquoteleft M$\\$\lbi{x}{V}{R}=R[V/x]$  &
$\toin{(\toin{M}{x}{N} )}{y}{N'}\stackrel{\#x}{=}\toin{M}{x}{(\toin{N}{y}{N'})}$\\
$\toin{M}{x}{\lambda_i N_i}=\lambda_i (\toin{M}{x}{N_i})$ & 
$\toin{M}{x}{\lambda_y N}\stackrel{\#y}{=}\lambda_y(\toin{M}{x}{N})$
\end{tabular}
}}
}
\caption{\label{fig:eqs} Equations of dCBPV. These should be read as equations of typed terms: we impose them if we can derive that both sides of the equation are terms of the same type (in context). We write $\stackrel{\#x_1,\ldots,x_n}{=}$ to indicate that for the equation to hold, the identifiers $x_1,\ldots,x_n$ should, in both terms being equated, be replaced by fresh identifiers, in order to avoid unwanted variable bindings.\vspace{-5pt}}
\end{figure}

Figures \ref{fig:depcbvtrans} and \ref{fig:depcbntrans} show the natural candidate CBV- and CBN-translations of DTT with some of the usual connectives into dCBPV:  we treat sums, projection products, pattern matching dependent products and unit types,  dependent function types and identity types. We can define the translations for  projection dependent products as we have for their simple relatives, if we add projection dependent  products to dCBPV as is sketched in section \ref{sec:conn}.

It turns out that without Kleisli extensions for dependent functions, the CBV-translation is not well-defined as it results in untypable terms. The CBN-translation is, but only if we restrict to the weak (non-dependent) elimination rules for $\Sigma_{1\leq i\leq n}$-, $1$-, $\Sigma$- and $\Id$-types, meaning that the type we are eliminating into does not depend on the type being eliminated from. One would expect the CBV-translation to factorise as a translation into a dependently typed version of Moggi's monadic metalanguage without dependent Kleisli extensions, followed by a translation of this into dCBPV-. While the latter is definable, the former is ill-defined\footnote{Although, this may remind the reader of the situation of dependently typed dual intuitionistic linear logic (a comonadic language), where the Girard translation (essentially the CBN-translation) from DTT fails, we  appear to be observing a qualitatively different phenomenon rather than a mere mirror image.}.  Perhaps this is a (partial) explanation of why all (CBV) dependently typed languages with effects have encapsulated the effects in a monad. The exceptions are non-termination and recursion. As we shall see in section \ref{sec:op}, dependent Kleisli extensions are  straightforward in that case without imposing subtyping conditions, which means we can treat these as a first class effects and do not have to encapsulate them in a~monad.

It seems  likely that one could obtain soundness and completeness results for these translations with respect to the correct equational theories for CBV- and CBN-DTT. As we are not aware of any such equational theories being described in literature, we propose to  \emph{define} these through their translations into dCBPV. Both  translations might result in a broken $\eta$-law for $\Id$-types, in presence of effects, even if we assume one in dCBPV.

\begin{figure}[!bt]
\fbox{\resizebox{\linewidth}{!}{
\parbox{1.17\linewidth}{
\begin{tabular}{l|l||l|l}
\textbf{CBV type}  & \textbf{CBPV type} & \textbf{CBV term } & \textbf{CBPV term}\\
\hline
$\Gamma\vdash A\type$ & $\vect{UF}\Gamma^v\vdash A^v\vtype$ & $x_1:A_1,\ldots,x_m:A_m$ & $x_1:A_1^v,\ldots,x_m:A_m^v[\ldots \tr x_{i}/z_{i}\ldots] $\\
 & & $\vdash M:A$ & $\vdash^c M^v:F(A^v[\tr x_1/z_1,\ldots,\tr x_n/z_n])$\hspace{0pt}\;\\
 && $x$& $\return x$\\
  &&$\lbi{x}{M}{N}$ & $\toin{M^v}{x}{N^v}$ \\
$\Sigma_{1\leq i\leq n }A_i$ & $\Sigma_{1\leq i\leq n }A_i^v$& $\langle i,M\rangle $&$\toin{M^v}{x}{\return \langle i, x\rangle }$ \\
 &&$\sipm{M}{i}{x}{N_i}$& $\toin{M^v}{z}{(\sipm{z}{i}{x}{N_i^v})}$\\
$\Pi_{1\leq i\leq n}A_i $ & $U\Pi_{1\leq i \leq n} FA_i^v$ &$\lambda_iM_i$ &$\return \thunk (\lambda_i  M_i^v)$\\
&&$i\textquoteleft  N $&$\toin{N^v}{z}{(i\textquoteleft \force z)}$\\
$\Pi_{x:A} A'$ & $U(\Pi_{x:A^v} F A'^v[\tr x/z])$\hspace{0pt}\; & $\lambda_x M$&$\return \thunk \lambda_x M^v$\\
&&$M\textquoteleft N$ &$\toin{M^v}{x}{(\toin{N^v}{z}{(x\textquoteleft \force z)})}$\\
$1$ & $1$ & $\langle\rangle$ & $ \return \langle\rangle$  \\
&&$\upm{M}{N}$&$\toin{M^v}{z}{(\upm{z}{N^v})}$\\
$\Sigma_{x:A}  A'$ & $\Sigma_{x:A^v} A'^v[\tr x/z]$ & $ \langle M, N\rangle $  & $\toin{M^v}{x}{(\toin{N^v}{y}{\return \langle x,y\rangle})}$\\
&&$\sipm{M}{x}{y}{N}$&$\toin{M^v}{z}{(\sipm{z}{x}{y}{N^v})}$\\
$\Id_A(M,N)$& {$\Id_{ UF A^v}( \thunk M^v$,} &$\refl M$& $\toin{M^v}{z}{\return \refl \tr z} $ \\
&{$ \thunk N^v)$}&$\idpm{M}{x}{N}$\hspace{30pt}\;& $\toin{M^v}{z}{(\idpm{z}{y}{}}$\\
&&& $(\toin{\force y}{x}{N^v}))$
\end{tabular}}}
}
\caption{\label{fig:depcbvtrans} A translation of dependently typed CBV into dCBPV. We write $\tr$ as an abbreviation for $\thunk\return$ and $\vect{UF}\Gamma:=z_1:UF A_1,\ldots,z_n:UFA_n$ for a context $\Gamma=x_1:A_1,\ldots,x_n:A_n$.\vspace{0pt}}
\end{figure}
\begin{figure}[!ht]
\fbox{\resizebox{\linewidth}{!}{
\parbox{1.17\linewidth}{
\begin{tabular}{l|l||l|l}
\textbf{CBN type}  & \textbf{CBPV type} & \textbf{CBN term } & \textbf{CBPV term }\\
\hline
${\Gamma}\vdash\ct{B}\type$ &  $\vect{U}{\Gamma^n}\vdash\ct{B}^n\ctype$& $x_1:\ct{B}_1,\ldots,x_m:\ct{B}_m\vdash M:\ct{B}$& $x_1:U\ct{B}_1^n,\ldots,x_m:U\ct{B}_m^n\vdash^c M^n:\ct{B}^n$ \\
 && $x$& $\force x$\\
  & & $\lbi{x}{M}{N}$ & $\lbi{x}{(\thunk M^n)}{N^n}$ \\
$\Sigma_{1\leq i\leq n }\ct{B}_i$ & $F\Sigma_{1\leq i\leq n }U\ct{B}_i^n$& $\langle i,M\rangle $&$\return \langle i,\thunk M^n\rangle $ \\
 & &$\sipm{M}{i}{x}{N_i}$&$\toin{M^n}{z}{(\sipm{z}{i}{x}{N_i^n})}$ \\
$\Pi_{1\leq i\leq n}\ct{B}_i $ & $\Pi_{1\leq i \leq n} \ct{B}_i^n$ & $\lambda_iM_i$& $\lambda_iM_i^n$\\
 && $i\textquoteleft M$ & $i\textquoteleft M^n$\\
$\Pi_{x:\ct{B}} \ct{B'}$ & $\Pi_{x:U\ct{B}^n} \ct{B'}^n$ & $\lambda_x M $& $\lambda_xM^n$\\
 &&$N\textquoteleft M$ & $(\thunk N^n) \textquoteleft M^n$ \\
$1$ & $F1$ & $\langle\rangle$ & $\return \langle\rangle$  \\
&&$\upm{M}{N}$&$\toin{M^n}{z}{(\upm{z}{N^n})}$\\
$\Sigma_{x:\ct{B}} \ct{B'}$ & $F(\Sigma_{x:U\ct{B}^n}  U\ct{B'}^n)$ & $\langle M, N\rangle $  & $\return \langle \thunk M^n,\thunk N^n\rangle$\\
&& $\sipm{M}{x}{y}{N}$& $\toin{M^n}{z}{(\sipm{z}{x}{y}{N^n})}$ \\
$\Id_{\ct{B}}(M,M')$&$F(\Id_{U\ct{B}}(\thunk M^n,$\quad\; & $\refl M$& $\return \refl \thunk M^n$\\
&$\thunk M'{}^n))$&$\idpm{M}{x}{N}$& $\toin{M^n}{z}{(\idpm{z}{x}{N^n})}\quad\;$
\end{tabular}
}}
}
\caption{\label{fig:depcbntrans} A translation of dependently typed CBN into dCBPV. We write $\vect{U}\Gamma:=x_1:U A_1,\ldots,x_n:UA_n$ for a context $\Gamma=x_1:A_1,\ldots,x_n:A_n$.\vspace{-10pt}}
\end{figure}
\nopagebreak

\section{Abstract Categorical Semantics}\label{sec:sem}
\vspace{-5pt}
We have reached the point in the story that was our initial motivation to study dCBPV: its very natural categorical semantics.  To formulate our dCBPV models, we recall the notion of an indexed category with full and faithful democratic comprehension, which is equivalent to Jacobs' notion of a split full democratic comprehension category with unit \cite{jacobs1993comprehension}.
\begin{definition}[Comprehension Axiom] Let $\Bcat^{op}\ra{\Ccat}\Cat$ be an indexed category (a functor to the category $\Cat$ of small categories). Given $B'\ra{f} B$ in $\Bcat$, let us write the change of base functor $\Ccat(B)\ra{}\Ccat(B')$ as $-\{f\}$. Recall that $\Ccat$ satisfies the \emph{comprehension axiom} if
\begin{itemize}
\item $\Bcat$ has a terminal object $\cdot$ and all fibres $\Ccat(B)$ have terminal objects $1_B$ that are stable under change of base;
\item the presheaves \mbox{
\begin{diagram}
(B'\ra{f}B) & \rMapsto  & \Ccat(B')(1,C\{f\})
 \end{diagram}} on $\Bcat/B$
are representable: we have representing objects $B.C\ra{\proj{B}{C}}B$ and natural bijections $
\Ccat(B')(1,C\{f\})\ra{\langle f,-\rangle } \Bcat/B(f,\proj{B}{C}) $.
\end{itemize}
Write $\diagv{B}{C}$ for the (universal) element of $\Ccat(B.C)(1,C\{\proj{B}{C}\})$ corresponding to $\id_{\proj{B}{C}}$. Define  \mbox{\small\begin{diagram} B.C & \rTo^{\diag{B}{C}:=\langle\id_{B.C},\diagv{B}{C} \rangle} & B.C.C\{\proj{B}{C}\}\end{diagram}},  \mbox{\small\begin{diagram}B'.C\{f\} & \rTo^{\mathbf{q}_{f,C}:=\langle\proj{B'}{C\{f\}};f,\diagv{B'}{C\{f\}}\rangle} & B.C\end{diagram}} and \linebreak
\mbox{\small\begin{diagram}
\Ccat(B)(C',C)& \rTo^{\proj{B}{-}:=\langle \proj{B}{C'},\diagv{B}{C'};-\{\proj{B}{C'}\}\rangle} &\Bcat/B(\proj{B}{C'},\proj{B}{C}).
\end{diagram}} 
When these last maps $\proj{B}{-}$  are full and faithful for all $B$, we call the comprehension \emph{full and faithful}, respectively. When the comprehension induces an equivalence $\Ccat(\cdot)\cong \Bcat$, we call it \emph{democratic}.
\end{definition}

\begin{definition}[dCBPV- Model] By a \emph{dCBPV- model}, we shall mean the following data.
\begin{itemize}
\item An indexed category $\Bcat^{op}\ra{\Ccat}\Cat$ with full and faithful democratic comprehension;
\item An indexed category $\Bcat^{op}\ra{\Dcat}\Cat$;
\item An indexed adjunction $F\dashv U: \Dcat \leftrightarrows \Ccat$ (adjunctions compatible with change of base);
\item $\Pi$-types in $\Dcat$: right adjoints $-\{\proj{A}{B}\}\dashv \Pi_B:\Dcat(A.B)\leftrightarrows\Dcat(A)$ satisfying the right Beck-Chevalley condition (a technical condition, compatibility with substitution \cite{jacobs1993comprehension});
\item Finite products ($\Pi_{1\leq i\leq n} D_i$) in the fibres of $\Dcat$, stable under change of base;
\item $\Sigma$-types in $\Ccat$: objects $\Sigma_CD$ of $\Ccat(B)$ such that $\proj{B}{\Sigma_CD}=\proj{B.C}{D};\proj{B}{C}$;
\item $\Id$-types in $\Ccat$: objects $\Id_C$ of $\Ccat(B.C.C)$ such that $\proj{B.C.C}{\Id_C}=\diag{B}{C}$;
\item $0,+$-types in $\Ccat$: finite coproducts ($\Sigma_{1\leq i\leq n} C_i$) in the fibres of $\Ccat$, stable under change of base, such that the following canonical morphisms are bijections,  for $\mathcal{E}\in\{\Ccat,\Dcat\}$,\\
$\mathcal{E}(C.\Sigma_{1\leq i\leq n}C_i)(X,Y)\ra{}\Pi_{1\leq i\leq n}\mathcal{E}(C.C_i)(X\{\proj{C}{\langle i,\id_{C_i}\rangle } \},Y\{\proj{C}{\langle i,\id_{C_i}\rangle }\})$;
\end{itemize}
\end{definition}
\begin{definition}[dCBPV+ Model] By a \emph{dCBPV+ model}, we mean a dCBPV- model with specified maps $\Ccat(\Gamma.A.\Gamma'\{\proj{\Gamma}{\eta_A }\})(1,UB\{\mathbf{q}_{\proj{\Gamma}{\eta_A},\Gamma'}\})\ra{(-)^*}\Ccat(\Gamma.UFA.\Gamma')(1,UB),$ called \emph{dependent Kleisli extensions},
where   $\eta$ is the unit of $F\dashv U$, such that $(-)^*$  agrees with the usual Kleisli extensions of $F\dashv U$ and is compatible with $\eta$ and $-\{-\}$ in the obvious sense.\end{definition}
\begin{remark}
Note that, for both notion of model, we retrieve (up to equivalence) Levy's adjunction models for CBPV if we restrict to $\Ccat=\self(\Bcat)$, where $\self(\Bcat)(B)(B',B''):=\Bcat(B\times B',B'')$, and $\Dcat$ that are locally $\Bcat$-indexed in the sense that change of base is the identity on objects and we drop the requirement of identity types (which then corresponds to objects $1/B$ such that $1/B\times B\cong 1$). In particular, stability of the adjunction under change of base then implies the strength of the monad $T=UF$ and stability of coproducts becomes distributivity. More generally, in a dCBPV- model, we can define a strength $\Sigma_ATB\ra{s_{A,B}}T\Sigma_AB$, while a costrength $\Sigma_{TA}B\ra{s_{A,B}'}T\Sigma_{A} TB\{\proj{\Gamma}{\eta_A }\} $ (hence a pairing $\Sigma_{TA} {TB}\ra{t_{A,B}} T\Sigma_AB\{\proj{\Gamma}{\eta_A }\}$) can only be defined in dCBPV+ models.
\end{remark}
This semantics is sound and complete.
\begin{theorem}[dCBPV Semantics] We have a sound interpretation of dCBPV- in a dCBPV- model and of dCBPV+ in a dCBPV+ model, where we also give the interpretation of the stack judgement of section \ref{sec:op} (together with the obvious interpretation of terms, e.g. interpreting $\toin{M}{x}{N}$ using Kleisli extensions, which we leave to the imagination of the reader):\\\\\resizebox{1.0\linewidth}{!}{\hspace{-10pt}\begin{tabular}{lll}\begin{tabular}{l}
$\sem{\cdot} = \cdot$\\
$\sem{\Gamma,x:A} = \sem{\Gamma}.\sem{A}$\\
$\sem{\Gamma\vdash^v A}=\Ccat(\sem{\Gamma})(1,\sem{A})$\\
$\sem{\Gamma\vdash^c \ct{B}}  = \Dcat(\sem{\Gamma})(F1,\sem{\ct{B}})$\\
$\sem{\Gamma;\ct{B}\vdash^k \ct{C}} = \Dcat(\sem{\Gamma})(\sem{\ct{B}},\sem{\ct{C}})$
\end{tabular}&\begin{tabular}{l}
$\sem{A[V/x]}  = \sem{A}\{\langle \langle \id_{\sem{\Gamma}},\sem{V}\rangle,\id_{\sem{\Gamma'[V/x]}}\rangle\}$\\$
\sem{\ct{B}[V/x]} = \sem{\ct{B}}\{\langle \langle \id_{\sem{\Gamma}},\sem{V}\rangle,\id_{\sem{\Gamma'[V/x]}}\rangle \}$\\$\sem{\Id_A(V,V')}  = \Id_{\sem{A}}\{\langle\langle  \id_{\sem{\Gamma}} , \sem{V}\rangle ,\sem{V'}\rangle \}$\\$
\sem{\Sigma_{1\leq i\leq n}A_i}=\Sigma_{1\leq i\leq n}\sem{A_i}$\\$
\sem{\Sigma_{x:A} A'} = \Sigma_{\sem{A}} \sem{A'}$\end{tabular}
&\begin{tabular}{l}
$
\sem{1}=1$\\$
\sem{\Pi_{1\leq i\leq n}\ct{B}_i}=\Pi_{1\leq i\leq n}\sem{\ct{B}_i}$\\$
\sem{\Pi_{x:A} \ct{B}}  = \Pi_{\sem{A}}\sem{\ct{B}}$\\$
\sem{FA}=F\sem{A}$\\$
\sem{U\ct{B}} = U\sem{\ct{B}}.$\end{tabular}
\end{tabular}}
\;\\
The interpretations in such categories are complete: an equality of values or computations holds in all interpretations iff it is provable in the syntax of dCBPV. In fact, if we add the obvious (admissible) weakening and exchange rules to dCBPV,  the interpretation defines an onto correspondence between categorical models and syntactic theories in dCBPV which satisfy mutual soundness and completeness results. This correspondence becomes 1-1 and we obtain completeness for the stack judgement if we include complex stacks.
\end{theorem}
\begin{proof}[Proof (sketch)] The proof goes almost entirely along the lines of the soundness and completeness proofs for linear DTT in \cite{vakar2014syntax}. For  completeness result, we build a syntactic category, after conservatively extending our syntax with complex stacks as in \cite{levy2005adjunction}. \end{proof}
\vspace{-10pt}
\section{Concrete Models}\label{sec:mod}
\vspace{-5pt}
We can first note that if we restrict to the identity adjunction, both dCBPV- and dCBPV+ reduce to a reformulation of Jacobs' notion of a full split democratic comprehension category with $\Sigma$-, $\Sigma_{1\leq i\leq n}$-, $\Pi$- and extensional $\Id$-types, which is a standard notion of model of pure DTT \cite{jacobs1993comprehension}. An example is the usual families of sets model $\Fam(\Set)$ of pure DTT. (Recall that $\Fam(\Set)$ is defined as the restriction to $\Set^{op}\subseteq \Cat^{op}$ of the $\Cat$-enriched hom-functor into $\Set$.) In particular, this shows consistency of the calculi.
\begin{theorem}
Both dCBPV- and dCBPV+ are consistent.
\end{theorem}
More interestingly, any model of linear DTT supporting the appropriate connectives \cite{vakar2014syntax,vakar2015syntax} gives rise to a model of dCBPV-, modelling commutative effects.
\begin{theorem}The notion of a model given by \cite{vakar2014syntax} for the dependently typed linear-non-linear logic of \cite{krishnaswami2015integrating} with the additional connectives of finite additive disjunctions is precisely a dCBPV- model such that we have symmetric monoidal closed structures on the fibres of $\Dcat$, stable under change of base, s.t. $F$ consists of strong monoidal functors (sending nullary and binary products in $\Ccat$ to $I$ and $\otimes$ in $\Dcat$) and which supports $\Sigma_{F-}^\otimes$-types (see section \ref{sec:conn}).
\end{theorem}
By analogy with the simply typed case, models of  DTT on which we have an indexed monad are again a source of  dCBPV- models (indexing of the monad corresponding to a strength).
\begin{theorem}\label{thm:emmodels}
Let $\Bcat^{op}\ra{\Ccat}\Cat$ be a model of pure DTT (as above) on which we have an indexed monad $T$ (a family of monads, stable under change of base). Then, the indexed Eilenberg-Moore adjunction $\Ccat\leftrightarrows \Ccat^T$ gives a model of dCBPV-.
\end{theorem}
\begin{proof}[Proof (sketch)] As in the simply typed setting, a product of algebras is just the product of their carriers equipped with the obvious algebra structure. Indeed, it is a basic fact from category theory that the forgetful functor from the Eilenberg-Moore category creates limits. Given an object $TB\ra{k}B$ of $\Ccat^T(\Gamma.A)$, we note that we also obtain a canonical $T$-algebra structure on $\Pi$-types of carriers (starting from the identity on $\Pi_A B$), our definition of $\Pi_{A}k$:\vspace{-5pt}\\\resizebox{\linewidth}{!}{\hspace{-10pt}\parbox{1.05\linewidth}{
\begin{align*}
\Ccat(\Gamma)(\Pi_AB,\Pi_A B)&\cong \Ccat(\Gamma.A)((\Pi_AB)\{\proj{\Gamma}{A}\},B)
\ra{T}  \Ccat(\Gamma.A)(T((\Pi_AB)\{\proj{\Gamma}{A}\}),TB)\\
&\cong  \Ccat(\Gamma.A)((T\Pi_AB)\{\proj{\Gamma}{A}\},TB)\ra{-;k}  \Ccat(\Gamma.A)((T\Pi_AB)\{\proj{\Gamma}{A}\},B)\cong \Ccat(\Gamma)(T\Pi_AB,\Pi_AB).\vspace{-5pt}
\end{align*}}}
We leave the verification of the $T$-algebra axioms to the reader. 
\end{proof}

A concrete example to which we can apply the previous theorem is obtained for any monad $T$ on $\Set$. Indeed, we can lift $T$ (pointwise) to an indexed monad on $\Fam(\Set)$. In a different vein, given a model $\Ccat$ of pure DTT, the usual exception, global state, reader, writer and continuation monads (which we form using objects of $\Ccat(\cdot)$) give rise to indexed monads, hence we obtain models of dCBPV-. More exotic examples are the many indexed monads that arise from homotopy type theory, like truncation  and cohesion (shape and sharp) modalities \cite{hottbook,shulman2015brouwer,schreiber2014quantum}. A caveat there is that the identity types  are intensional and that many equations are often only assumed up to propositional rather than judgemental~equality.

Models of dCBPV+ are harder to come by. Not every dCBPV- model generated by theorem \ref{thm:emmodels} admits dependent Kleisli extensions. Their existence needs to be assessed on a case-by-case basis. Moreover, one  indexed monad might be given dependent Kleisli extensions in several inequivalent ways. Therefore, we treat some dCBPV- models for common effects and discuss the (im)possibility of dependent Kleisli extensions.

Fortunately, the exceptions (and divergence) monads $(-)+E$ on $\Set$ admit~dependent Kleisli extensions, hence give rise to models of dCBPV+. Indeed, for a dependent function $f\in \Pi_{a\in A}B(a)+E$, we can define $f^*\in \Pi_{x\in A+E}B(x)+E$ by setting $f^*(a):=f(a)$ for $a\in A$ and else $f^*(e):=e$. (Generally, it suffices to give dependent Kleisli extensions for $B=FA'$.)

Similarly, the domain semantics of DTT with recursion of \cite{palmgren1990domain} yields a model of dCBPV+ with intensional $\Id$-types. $\Ccat$ is defined to be an indexed category of continuous families of Scott predomains (disjoint unions of Scott domains) with continuous functions. $\Sigma$-types are interpreted as the usual set theoretic way and are endowed with the product order, $\Sigma_{1\leq i\leq n}$-types are disjoint unions and $\Id_A(x,y):= \{z\in A\;|\; z\leq x,y\}$ with the induced order from $A$. $\Dcat$ is defined to be the  indexed category over the category of Scott predomains of continuous families of Scott domains with strict continuous functions. $U$ is defined to be the usual inclusion, while $F$ is the lift. $\Pi_A \ct{B}$ is defined to be the set of continuous sections of $\Sigma_A U\ct{B}\ra{\proj{A}{U\ct{B}}}A$ with the pointwise order. We note that for $f\in \Pi_{A}U\ct{B}$, we can define $f^*(a):=f(a)$ for $a\in A$ and $f^*(\bot)=\bot$.

By contrast, a non-trivial writer monad $M\times -$ (for a monoid $M$) on $\Set$ does not admit dependent Kleisli extensions. Indeed, we could take $B$ to be a predicate on $M\times A$ such that $B(1_M,a)=\{*\}$ and otherwise $B(m,a)=\emptyset$ (which expresses that nothing is being written). $*\in \Pi_{a\in A}B(1_m,a)$ does not have a dependent Kleisli extension unless $A=\emptyset$.

Similarly, a non-trivial reader monad $(-)^S$ on $\Set$ does not admit dependent Kleisli extensions. Indeed, we can define predicates $B$ to specify that all $s\in S$ yield the same value. An analogous argument applies to non-trivial global state monads $(S\times -)^S$ on $\Set$.

For a continuation monad $R^{(R^-)}$ on $\Set$, one would hope to define dependent Kleisli extensions for $f\in \Pi_{a\in A}R^{(R^{B(\mathsf{ev}_a)})}$ as $f^*(t)(k):=t(\lambda_a f(a)(k))$. However, this is only well-defined if we have $\forall_{a\in A(c)}R^{B(c,t)}\subseteq R^{B(c,\mathsf{ev}_a)}$. In particular, we would have that $B(c,\mathsf{ev}_A)=B(c,\mathsf{ev}_{A'})$ for all $a,a'\in A(c)$.

Somewhat dubiously, we can define dependent Kleisli extensions for the erratic choice monad $\mathcal{P}$ on $\Set$ as $f^*(t):=(\bigcup_{a\in t} f(a))\cap B(t)$, for $f\in \Pi_{a\in A}\mathcal{P}B(a)$. This might not correspond to the desired operational behaviour: indeed, non-deterministic branching is not accumulated if it contradicts typing. 

Many of the non-examples above of dCBPV+ models, could be mended by restricting the class of predicates we consider, which would involve passing to a less simplistic model of dependent types than mere families of sets. For example, for the writer monad, we want to demand for types $B$ depending on $M\times A$ the inclusion $B(1_M,a)\subseteq B(m,a)$. We will see this idea reflected in the subject reduction property of the operational semantics of dCBPV+.
\vspace{-5pt}
\section{Operational Semantics}\label{sec:op}
\vspace{-5pt}
Importantly, CBPV admits a natural operational semantics that, for ground terms, reproduces the usual operational semantics of CBV and CBN under the specified translations into CBPV \cite{levy2012call}. This can easily be extended to dCBPV.

We define the operational semantics on dCBPV terms not involving \emph{complex values} which unnecessarily complicate the presentation of the operational semantics. Complex values are defined to be values containing $\mathsf{pm}\;\;\mathsf{as}\;\;\mathsf{in}\;\;$- and $\lbi{}{}{}$-constructs. As the normalization of  values do not produce effects, all reasonable evaluation strategies for them are observationally indistinguishable and we could choose our favourite. However, we may still want to allow complex values to occur in types, as this greatly increases the expressivity of our type system.

To make sure we lose no expressive power while expressive power when excluding complex values, we add the rule of figure \ref{fig:dcbpvextrarule} to dCBPV. We note that this rule introduces no new terms to dCBPV, at least up to judgemental equality, as $\toin{M}{x}{N}=\toin{\return{V}}{x}{N}=\lbi{x}{V}{N}$. This means that it is invisible to the categorical semantics. It allows us to define the operational semantics for a wider range of terms, however. It allows us to  eliminate complex values from computations (even at types built with complex values!).

\begin{figure}[!ht]\fbox{\resizebox{\linewidth}{!}{\parbox{1.22\linewidth}{
\parbox{\linewidth}{\;\hspace{-1pt}\;
\AxiomC{$\Gamma\vdash^c M= \return{V}: FA$}
\AxiomC{$\Gamma,x:A,\Gamma'\vdash^c N:\ct{B} $}
\BinaryInfC{$\Gamma,\Gamma'[V/x]\vdash^c \toin{M}{x}{N}:\ct{B}[V/x]$}
\DisplayProof\hspace{400pt}\;
}}}}
\caption{\label{fig:dcbpvextrarule} An extra rule to add to dCBPV which does not introduce any new terms.}
\end{figure}
\vspace{-10pt}
\begin{theorem}[Redundancy of Complex Values for Computations] For any dCBPV- or dCBPV+ computation $\Gamma\vdash^c M:\ct{B}$, there is a computation $\Gamma\vdash^c \widetilde{M}:\ct{B}$ (of dCBPV- or dCBPV+, respectively) which does not contain complex values, such that $\Gamma\vdash^c M=\widetilde{M}:\ct{B}$. Moreover, both the CBV- and CBN-translations only produce complex-value-free computations.\end{theorem}
We present a small-step operational semantics for (complex value-free) dCBPV computations in terms of a simple abstract machine that Levy calls the CK-machine. The configuration of such a machine consists of a pair $M,K$ where $\Gamma\vdash^c M:\ct{B}$ is a \emph{complex value-free computation} and $\Gamma;\ct{B}\vdash^k K:\ct{C}$ is a compatible (simple) \emph{stack}. We call $\ct{C}$ the type of the configuration. Stacks\footnote{To be precise, these are what Levy calls simple stacks  \cite{levy2012call}. Analogous to complex values, one can also conservatively extend the calculus with so-called \emph{complex stacks}, e.g. by allowing pattern matching into stacks. This gives us a 1-1 correspondence between categorical models and syntactic theories.} are formed according to the rules of figure \ref{fig:simplestacks}.

\begin{figure}[!ht]
\fbox{\resizebox{\linewidth}{!}{\hspace{-9pt}
\parbox{1.23\linewidth}{
\begin{tabular}{llll}
\AxiomC{$\vdash\Gamma;\ct{C}\ctxt$}
\UnaryInfC{$\Gamma;\ct{C}\vdash^k \nil : \ct{C}$}
\DisplayProof\hspace{-5pt}
& 
\AxiomC{$\Gamma,x:A\vdash^c M:\ct{B}$\hspace{-12pt}}
\AxiomC{$\Gamma;\ct{B}\vdash^k K:\ct{C}$}
\BinaryInfC{$\Gamma;FA\vdash^k \toin{[\cdot]}{x}{M}::K:\ct{C}$}
\DisplayProof\hspace{-5pt}
&
\AxiomC{$\Gamma;\ct{B_j}\vdash^k K:\ct{C}$}
\UnaryInfC{$\Gamma;\Pi_{1\leq i\leq n}\ct{B_i}\vdash^k j::K:\ct{C}$}
\DisplayProof\hspace{-5pt}
&
\AxiomC{$\Gamma\vdash^v V:A$\hspace{-12pt}}
\AxiomC{$\Gamma;\ct{B}[V/x]\vdash^k K:\ct{C}$}
\BinaryInfC{$\Gamma;\Pi_{x:A} \ct{B}\vdash^k V::K:\ct{C}$}
\DisplayProof
\end{tabular}
}
}
}
\caption{\label{fig:simplestacks} The rules for forming (simple) stacks.}
\end{figure}\vspace{-5pt}
The initial configurations, transitions (which embody directed versions of the $\beta$-rules of our equational theory) and terminal configurations in the evaluation of a computation $\Gamma\vdash^c M:\ct{C}$ on the CK-machine are specified by figure \ref{fig:ckmachine}.
\begin{figure}[!ht]
\fbox{\resizebox{\linewidth}{!}{
\parbox{1.24\linewidth}{
\begin{tabular}{ll}
\parbox{.245\linewidth}{\hspace{-12pt}
\begin{tabular}{lllllll}
\textbf{Initial Configuration}\hspace{-100pt}\;&&&\\
$M$ &,& $\nil$ &
&&&\\
&&&\\
&&&\\
\textbf{Terminal Configurations}\hspace{-100pt}\;&&\\
$\return V$ &,& $\nil$\\
$\lambda_i M_i$ &,& $\nil$\\
$\lambda_x M$ &,& $\nil$\\
$\force z$ &,& $ K$\\
$\sipm{z}{i}{x}{M_i}$\hspace{-5pt}\; &,& $K$\\
$\upm{z}{M}$ &,& $K$\\
$\sipm{z}{x}{y}{M}$\hspace{-5pt}\; &,& $K$\\
$\idpm{z}{x}{M}$\hspace{-7pt}\;  &,& $K$
\end{tabular}}&
\parbox{.5\linewidth}{
\begin{tabular}{lllllll}
\textbf{Transitions}\hspace{-100pt}&&&&&&\\
$\lbi{V}{x}{M}$ &,\hspace{-5pt}\;& $K$& $\leadsto$ & $M[V/x]$ &,\hspace{-5pt}\;& $K$ \\
$ \toin{M}{x}{N} $ &,\hspace{-5pt}\;& $K$& $\leadsto$ & $ M$ &,\hspace{-5pt}\;& $\toin{[\cdot]}{x}{N}::K$ \\
$   \return V     $ \hspace{20pt}&,\hspace{-5pt}\;& $\toin{[\cdot]}{x}{N}::K$\hspace{-6pt}\;& $\leadsto$ & $   N[V/x]       $ &,\hspace{-5pt}\;& $K$ \\
$   \force\thunk M     $ &,\hspace{-5pt}\;& $K$& $\leadsto$ & $   M       $ &,\hspace{-5pt}\;& $K$ \\
$ \sipm{\langle j,V\rangle}{i}{x}{M_i}  \hspace{-8pt}\;    $ &,\hspace{-5pt}\;& $K$& $\leadsto$ & $   M_j[V/x]       $ &,\hspace{-5pt}\;& $K$ \\
$ \upm{\langle\rangle}{M}       $ &,\hspace{-5pt}\;& $K$& $\leadsto$ & $ M         $ &,\hspace{-5pt}\;& $K$ \\
$  \sipm{\langle V,W\rangle }{x}{y}{M} \hspace{-8pt}\;     $ &,\hspace{-5pt}\;& $K$& $\leadsto$ & $  M[V/x,W/y] \hspace{-8pt}\;       $ &,\hspace{-5pt}\;& $K$ \\
 $  \idpm{(\refl V)}{x}{}{M}      $\hspace{-12pt}\;  &,\hspace{-5pt}\;& $K$& $\leadsto$ & $  M[V/x]        $ &,\hspace{-5pt}\;& $K$ \\
$ j\textquoteleft M       $ &,\hspace{-5pt}\;& $K$& $\leadsto$ & $       M   $ &,\hspace{-5pt}\;& $j::K$ \\
$ \lambda_i M_i       $ &,\hspace{-5pt}\;& $j::K$& $\leadsto$ & $    M_j      $ &,\hspace{-5pt}\;& $K$ \\
$ V\textquoteleft M       $ &,\hspace{-5pt}\;& $K$& $\leadsto$ & $      M   $ &,\hspace{-5pt}\;& $V::K$ \\
$ \lambda_x M       $ &,\hspace{-5pt}\;& $V::K$& $\leadsto$ & $   M[V/x]       $ &,\hspace{-5pt}\;& $K$ 
\end{tabular}}
\end{tabular}
}}
}
\caption{\label{fig:ckmachine} The behaviour of the CK-machine in the evaluation of a computation $\Gamma\vdash^c M:\ct{C}$.}\vspace{-10pt}
\end{figure}

The operational semantics of dCBPV- and dCBPV+ satisfy the following basic properties.
\begin{theorem}[Pure dCBPV Operational Properties]\label{thm:dcbpvopprop} For every configuration of the CK-machine, at most one transition applies. No transition applies precisely when the configuration is terminal. Every configuration of type $\ct{C}$ reduces, in a finite number of transitions, to a unique terminal configuration of type $\ct{C}$.
\end{theorem}
\begin{proof}[Proof (sketch)] The proof for dCBPV- is no different from the simply typed situation \cite{levy2012call}, as the transitions are defined on untyped configurations and are easily seen to preserve types. Note that types only depend on values, which never get reduced in our transitions.

Only the subject reduction property for dCBPV+ requires some thought. Suppose we start from a configuration $M,K$ with $\Gamma\vdash^c M:\ct{B}$ and $\Gamma;\ct{B}\vdash^k K:\ct{C}$. What could conceivably happen during a transition is that the resulting configuration $M',K'$ can be given types $\Gamma\vdash^c M':\ct{B}'$  and $\Gamma';\ct{B}''\vdash^k K':\ct{C}$, but not  $\Gamma\vdash \ct{B}'=\ct{B}''$.

The only transition that can possibly cause this problem is the one for initial computation $\return V$ (see figure \ref{fig:ckmachine}). Indeed, what could have happened is that we started from a configuration $\toin{M}{x}{N},K$, with $\Gamma\vdash^c \toin{M}{x}{N}: \ct{B}[\thunk M/z]$, which transitions to $M,\toin{[\cdot]}{x}{N}::K$. After this, we continue reducing $M$, until we end up in a configuration $\return V, \toin{[\cdot]}{x}{N}::K$. It is now critical when we apply one more transition (the dangerous one) that for the resulting configuration $N[V/x],K$ we again have that $\Gamma\vdash N[V/x]:\ct{B}[\thunk M/z]$, while we know that $\Gamma\vdash N[V/x]:\ct{B}[\tr V/z]$.

This is indeed the case in pure dCBPV+, as all minimal sequences of transitions of the shape $M,K\leadsto M',K$ just consist of an application of a directed version (left to right) of one of the judgemental equalities of figure \ref{fig:eqs}, hence $ M=\return V$ so $\Gamma\vdash N[V/x]:\ct{B}[\thunk M/z]$.
\end{proof}
\vspace{-10pt}
\section{Adding Effects}\label{sec:eff}
\vspace{-5pt}
We show by example how one adds effects to dCBPV, focussing on the operational semantics. Figure \ref{fig:effects} gives  examples of effects one could consider:  divergence, recursion, printing elements $m$ of some monoid $\mathcal{M}$, erratic choice from finitely many alternatives, errors $e$ from some set $E$, writing a global states $s\in S$ and reading a global state to $s$. The framework fits many more examples like probabilistic erratic choice, local references and control operators \cite{levy2012call}.
\begin{figure}[!ht]
\fbox{\resizebox{\linewidth}{!}{
\parbox{1.22\linewidth}{
\begin{tabular}{llll}
\AxiomC{}
\UnaryInfC{$\Gamma\vdash^c \diverge :\ct{B}$}
\DisplayProof\hspace{30pt}
&\AxiomC{$\Gamma,z:U\ct{B}\vdash^c M : \ct{B}$}
\UnaryInfC{$\Gamma\vdash^c \mu_z M : \ct{B}$}
\DisplayProof\hspace{30pt} &

\AxiomC{$\Gamma\vdash^c M:\ct{B}$}
\UnaryInfC{$\Gamma\vdash^c \print{m}M:\ct{B}$}
\DisplayProof\hspace{30pt}
&
\AxiomC{$\{\Gamma\vdash^c M_i:\ct{B}\}_{1\leq i\leq n}$}
\UnaryInfC{$\Gamma\vdash^c \nondet{i}{M_i}:\ct{B}$}
\DisplayProof  \\
\vspace{-5pt}&&&\\
\AxiomC{}
\UnaryInfC{$\Gamma\vdash^c \error{e} :\ct{B}$}
\DisplayProof
& &
\AxiomC{$\Gamma\vdash^c M:\ct{B}$}
\UnaryInfC{$\Gamma\vdash^c \writecell{s}M:\ct{B}$}
\DisplayProof
&
\AxiomC{$\{\Gamma\vdash^c M_s:\ct{B}\}_{s\in S}$}
\UnaryInfC{$\Gamma\vdash^c \readcell{s}{M_s}:\ct{B}$}
\DisplayProof

\end{tabular}
}
}}
\caption{\label{fig:effects} Some examples of effects we could add to CBPV.\vspace{-5pt}}
\end{figure}
\\For the operational semantics of printing and state, we need to add some hardware to our machine. Therefore, a configuration will now consist of a quadruple $M,K,m,s$ where $M,K$ are as before, $m$ is an element of a printing monoid $(\mathcal{M},\epsilon,*)$  modelling some channel for output and $s$ is an element of a finite pointed set of states $(S,s_0)$  holding the current value of our storage cell. We lift the operational semantics of all existing language constructs to this setting by specifying that they do not modify $m$ and $s$, that terminal configurations can have any value of $m$ and $s$ and that initial configurations always have value $m=\epsilon$ and $s=s_0$ for the fixed initial state $s_0$. Figure \ref{fig:opsemprint}  specifies the operational semantics for our effects.
 
\begin{figure}[!ht]
\fbox{
\resizebox{\linewidth}{!}{\hspace{-13pt}
\parbox{1.22\linewidth}{\begin{tabular}{ll}
\begin{tabular}{lllllllllllllll}
\textbf{Transitions}\hspace{-100pt}\;&&&&&&&&&&&&&&\\
$\diverge$ \hspace{-8pt}\; &,\hspace{-8pt}\; & $K$ \hspace{-8pt}\; &,\hspace{-8pt}\; & $m$\hspace{-8pt}\; &,\hspace{-8pt}\; &$s$ & $\leadsto$ & $\diverge$ \hspace{-8pt}\; &,\hspace{-8pt}\; & $K$  \hspace{-8pt}\; &,\hspace{-8pt}\; & $m$\hspace{-8pt}\; &,\hspace{-8pt}\; & $s$\\
$\mu_z M $ \hspace{-8pt}\; &,\hspace{-8pt}\; & $K$ \hspace{-8pt}\; &,\hspace{-8pt}\; & $m$\hspace{-8pt}\; &,\hspace{-8pt}\; &$s$ & $ \leadsto$ & $M[\thunk \mu_z M /z]$ \hspace{-8pt}\; &,\hspace{-8pt}\; & $K$  \hspace{-8pt}\; &,\hspace{-8pt}\; & $m$\hspace{-8pt}\; &,\hspace{-8pt}\; & $s$\\
$\nondet{i}{M_i}$ \hspace{-8pt}\; &,\hspace{-8pt}\; & $K$ \hspace{-8pt}\; &,\hspace{-8pt}\; & $m$\hspace{-8pt}\; &,\hspace{-8pt}\; &$s$ & $ \leadsto$ & $M_j$ \hspace{-8pt}\; &,\hspace{-8pt}\; & $K$\hspace{-8pt}\; &,\hspace{-8pt}\; & $m$\hspace{-8pt}\; &,\hspace{-8pt}\; & $s$\\
$\print n M$ \hspace{-8pt}\; &,\hspace{-8pt}\; & $K$\hspace{-8pt}\; &,\hspace{-8pt}\; & $m$& , & $s$ & $\leadsto$ & $M$ \hspace{-8pt}\; &,\hspace{-8pt}\; & $K$ & ,& $m*n$\hspace{-8pt}\; &,\hspace{-8pt}\; & $s$\\
$\writecell {s'} M$ \hspace{-8pt}\; &,\hspace{-8pt}\; & $K$\hspace{-8pt}\; &,\hspace{-8pt}\; & $m$& , & $s$ & $\leadsto$ & $M$ \hspace{-8pt}\; &,\hspace{-8pt}\; & $K$ \hspace{-8pt}\; &,\hspace{-8pt}\; & $m$\hspace{-8pt}\; &,\hspace{-8pt}\; & $s'$\\
$\readcell {s'} {M_{s'}}$ \hspace{-13pt}\;\hspace{-8pt}\; &,\hspace{-8pt}\; & $K$\hspace{-8pt}\; &,\hspace{-8pt}\; & $m$& , & $s$ & $\leadsto$ & $M_s$ \hspace{-8pt}\; &,\hspace{-8pt}\; & $K$ \hspace{-8pt}\; &,\hspace{-8pt}\; & $m$\hspace{-8pt}\; &,\hspace{-8pt}\; & $s$\\
\end{tabular}\hspace{-13pt}\;
&
\begin{tabular}{lllllll}
\textbf{Terminal Configurations}\hspace{-100pt}\; && && &&\\
$\error e$ \hspace{-10pt}\; & , \hspace{-8pt}\;& $K$ \hspace{-8pt}\;& ,\hspace{-8pt}\;& $m$\hspace{-8pt}\; & , \hspace{-8pt}\;& $s$ \\
&& && &&\\
&& && &&\\
&& && &&\\
&& && &&\\
&& && &&\\
\end{tabular}
\end{tabular}
}
}}
\caption{\label{fig:opsemprint} The operational semantics for divergence, recursion, erratic choice, errors, printing and writing and reading global state.\vspace{-10pt}}
\end{figure} 
We extend the results of the previous section to this effectful setting. While we will have to add some extra rules to dCBPV+, dCBPV- satisfies theorem \ref{thm:effectopprop} without any further rules.
\begin{theorem}[Effectful dCBPV Operational Properties]\label{thm:effectopprop} Transitions respect the type of the configuration. None applies precisely if we are in a terminal configuration. In absence of erratic choice, at most one transition applies to each configuration. In absence of divergence and recursion, every configuration reduces to a terminal one in a finite number of~steps.
\end{theorem}
As highlighted by the proof of theorem \ref{thm:dcbpvopprop}, to guarantee subject reduction for dCBPV+, we again need to establish that for every transition $M,K,m,s\leadsto M',K,m',s'$ of figure \ref{fig:effects}, $\Gamma\vdash^c N:\ct{B}[\thunk M'/z]$ implies that $\Gamma\vdash^c N:\ct{B}[\thunk M/z]$. A sufficient condition, would be to add equations $M=M'$. For divergence and errors, these equations would be vacuous, and for recursion, it gives us the usual equation $\mu_z M = M[\thunk\mu_zM/z]$. For printing, erratic choice and global state, however, these equations are not reasonable to demand. Instead, for these effects, we suggest to add explicitly the rules \resizebox{!}{12pt}{\mbox{\AxiomC{$\Gamma\vdash^c N:\ct{B}[\thunk M'/z]$}
\UnaryInfC{$\Gamma\vdash^c N:\ct{B}[\thunk ({M'})/z]$}
\DisplayProof}}.

Specifically, we see that a computation type in which we substitute (the thunk of) a computation with multiple branches (erratic choice, reading state) has to contain each of the types obtained by substituting any of the branches instead. Similarly, a type in which we substitute (the thunk of) a computation which writes (printing, writing a state) and then performs a computation $M$ has to contain the type obtained by just substituting $M$. In short: a type can only shrink as the computation unfolds. As anticipated, these rules restore the subject reduction property for effectful dCBPV+ and make theorem \ref{thm:effectopprop} hold for it too.

We can extend the CBV- and CBN-translations to the effectful setting following figure \ref{fig:transeff}. By analogy with the simply typed situation, this should induce the expected operational semantics, at least for terms of ground type, although we are not aware of such an operational semantics ever being outlined in literature. 
\begin{figure}[!ht]
\fbox{\resizebox{\linewidth}{!}{
\parbox{1.22\linewidth}{
\begin{tabular}{l|ll||ll|l}
\textbf{CBV Term} $M$ & \textbf{CBPV Term} $M^v$&\qquad\qquad\qquad\qquad && \textbf{CBN Term} $M$ & \textbf{CBPV Term} $M^n$\qquad\qquad\qquad\qquad\\
\hline
$\op(M_1,\ldots,M_m) $& $\op(M_1^v,\ldots, M_m^v)$ &&& $\op(M_1,\ldots,M_m) $& $\op(M_1^n,\ldots,M_m^n)$\\
$\mu_x M $ & $\mu_z (\toin{\force z}{x}{M^v})$ &&&$\mu_z M $ & $\mu_z M^n$
\end{tabular}}}}
\caption{\label{fig:transeff} The CBV- and CBN-translations for effectful terms. Here, we treat all algebraic effects (in particular, divergence, errors, erratic choice, printing and global state) uniformly by letting $\op$ range over their operations. $z$ is assumed to be fresh in the CBV-translation $\mu_x M$.\vspace{-10pt}}
\end{figure}
\begin{remark}[Type Checking] 
One could argue that a type checker is as important an operational aspect to the implementation of a DTT as a small-step semantics. We leave the description of a type checking algorithm to future work. We note that the core step in the implementation of a type checker is a normalization algorithm for directed versions (from left to right) of the judgemental equalities, as this would give us a normalization procedure for types. We hope to be able to construct such an algorithm for dCBPV- using normalization by evaluation by combining the techniques of \cite{abel2007normalization} and \cite{ahman2013normalization}. Our hope is that this will lead to a proof of decidable type checking of the system at least in absence of the $\eta$-law for $\Id$-types and without recursion. We note that the complexity of a type checking algorithm can vary widely depending on which equations we include for effects. The idea is that one only includes a basic set of program equations (perhaps including algebraicity equations for effects) as judgemental equalities to be able to decide type checking and to postulate other equations (like the Plotkin-Power equations for state) as propositional equalities, which can be used for manual or tactic-aided reasoning about effectful programs. Type checking for effectful dCBPV+ seems  more challenging, as the system essentially features a form of subtyping.
\end{remark}
\vspace{-5pt}
\section{More Connectives?}\label{sec:conn}
\vspace{-5pt}
We very briefly outline how to add more connectives to dCBPV.

A first class of connectives we consider are projection dependent products, or $\Sigma$-types on computations. Given a context $ \Gamma,z_1:U\ct{B_1},\ldots,z_n:U\ct{B_n} $, we can form these  types $\Gamma\vdash\Pi_{1\leq i\leq n}^{dep} \ct{B_i}\ctype$ and we construct and destruct their terms using the rules of figure \ref{fig:addsigma}.

\begin{figure}
[!ht]
\fbox{\resizebox{\linewidth}{!}{
\parbox{1.22\linewidth}{
\begin{tabular}{ll}

\AxiomC{$\{\Gamma\vdash^c M_i:\ct{B_i}[\thunk M_1/z_1,\ldots,\thunk M_{i-1}/z_{i-1}]\}_{1\leq i\leq n}$}
\UnaryInfC{$\Gamma \vdash^c \lambda_i M_i : \Pi_{1\leq i\leq n}^{dep}\ct{B_i}$}
\DisplayProof\qquad
&
\AxiomC{$\Gamma\vdash^c M:\Pi_{1\leq i \leq n}^{dep} \ct{B_i}$}
\UnaryInfC{$\Gamma\vdash^c i\textquoteleft M: \ct{B_i}[\thunk 1\textquoteleft M/z_1,\ldots,\thunk (i-1)\textquoteleft M/z_{i-1}]$}
\DisplayProof
\end{tabular}
}
}}
\caption{\label{fig:addsigma} Rules for projection dependent products, to which we add the obvious $\beta$- and $\eta$-laws.\vspace{0pt}}
\end{figure}
\noindent Their categorical semantics corresponds to having strong $n$-ary $\Sigma$-types in $\Dcat$ in the sense of objects $\Pi_{1\leq i\leq n}^{dep}\ct{B_i}$ such that  $\proj{\Gamma}{U\Pi_{1\leq i\leq n}^{dep}\ct{B_i}}=\proj{\Gamma.U\ct{B_1}.\cdots.U\ct{B_{n-1}}}{U\ct{B_n}};\ldots;\proj{\Gamma}{U\ct{B_1}}$. In particular, we see that $U\Pi_{1\leq i\leq 2}^{dep}\ct{B}_i \cong \Sigma_{U\ct{B}_1}U\ct{B}_2$. Similarly, we can define  equivalents of the other positive connectives $R$ like $\Sigma_{1\leq i\leq n}$- and $\Id$-types on computation types in the sense of connectives $R'(\ct{B}_1,\ldots,\ct{B}_n)$ such that $UR'(\ct{B}_1,\ldots,\ct{B}_n)\cong R(U\ct{B}_1,\ldots,U\ct{B}_n)$. These correspond to e.g.  additive $\Sigma$- and $\Id$-types in linear DTT. In all cases, we give an operational semantics where destructors push to the stack and constructors pop the stack and substitute.

Another connective from linear DTT that generalizes to dCBPV is the multiplicative $\Sigma$-type $\Gamma \vdash \Sigma_{FA}^\otimes\ct{B}\ctype$, for $\Gamma,x:A\vdash \ct{B}\ctype$. These type formers are dependently typed generalizations of the $F(-)\otimes -$-connective of the EEC (there written $!(-)\otimes-$). Their categorical semantics, as in linear DTT, is given by left adjoint functors $\Sigma_{FA}^\otimes \dashv -\{\proj{\Gamma}{A}\}$ to change of base in $\Dcat$, satisfying the left Beck-Chevalley~condition. In particular, they satisfy the equation $F\Sigma_{A}A'\cong\Sigma_{FA}FA'$.

Finally, to build a practically useful system, strong type forming mechanisms like inductive families and induction-recursion (including universes) should probably be added to dCBPV as constructions on value types. While it seems believable that we then still get a CBV- and CBN-translation in dCBPV+ for inductive families, the situation for universes is less straightforward. Another question is if such translations are desirable or if we are better off working with the simpler system dCBPV- in the first place.

\vspace{-5pt}
\section{Future Work}
\label{sec:concl}
\vspace{-5pt}
In this paper, we gave an extension to the realm of dependent types of Levy's CBPV analysis of CBV- and CBN-$\lambda$-calculi. We hope, one the one hand, that this can shed some light on the theoretical puzzle of how the relationship between effects and dependent types can be understood. On the other hand, we hope it can provide some guidance in the daunting but worthwhile challenge of combining dependent types with effects in a system which is simultaneously mathematically elegant and practically useful for writing certified real world software. To further achieve these two goals, we have the following future work in mind.

A first priority is the implementation of a type checker for dCBPV- (and perhaps dCBPV+). A second question to be addressed is what needs to be done to make dCBPV into a practically useful language for certified effectful programming. Does dCBPV- suffice for this or do we need to turn to dCBPV+? Thirdly, the recent CBN game semantics for DTT of \cite{abramsky2015games}, which can be extended to model recursion, local ground store and control operators, should give rise to a model of dCBPV+. Fourthly, we want to investigate if the CBV- and CBN-translations into dCBPV+ extend to more expressive type forming mechanisms like inductive families and inductive-recursive definitions like type universes. In particular, we hope this will lead to a better understanding of the rather enigmatic nature of CBV-type dependency. Finally, it remains to be seen what the status is of subject reduction in dCBPV+ in presence of other effects like control operators, local state and general references. We hope the game semantics of DTT can be of use here. 

\subparagraph*{Related Work} This paper is based on the preprint \cite{vakar2015framework}, which provides more context,~proofs and  discussion. Since this appeared, independent work by Ahman,~Ghani and Plotkin \cite{ahman2016depty} has been made public which partly overlaps with sections \ref{sec:syn} and \ref{sec:sem} of~this~paper. It describes a dependently typed  EEC, closely related to dCBPV- extended with complex stacks and $\Sigma_{F-}^\otimes$-types, and its categorical semantics which - modulo details - is a fibrational reformulation of our semantics for this calculus.  Additionally, our sections \ref{sec:syn} and \ref{sec:sem} consider a more expressive calculus dCBPV+ where we can substitute effectful computations in dependent functions. This allows us to define CBV- and CBN-translations into it. Also relevant is the work on linear  \cite{krishnaswami2015integrating,vakar2015syntax} and polarised \cite{licata2009positively} DTT and on domain and game semantics for DTT~\cite{palmgren1990domain,abramsky2015games}.

\subparagraph*{Acknowledgements}
I want to thank Tim Zakian, Alex Kavvos and Sam Staton for many interesting discussions and Paul Blain Levy for his explanations. I am grateful to Samson Abramsky for his support. The author is funded by the EPSRC and the Clarendon Fund.





\nocite{Simpson}

{
\bibliography{bibliography}}

\end{document}